\documentstyle[fleqn,12pt]{article}

\newcommand{\ra}{\rightarrow}
\newcommand{\bra}{\langle}
\newcommand{\ket}{\rangle}
\newcommand{\be}{\begin{equation}}
\newcommand{\ee}{\end{equation}}
\newcommand{\bea}{\begin{eqnarray}}
\newcommand{\eea}{\end{eqnarray}}
\newcommand{\eps}{\varepsilon}

\newcommand{\e}{\mbox{e}} 

\newcommand{\ffi}{\varphi}
\newcommand{\ode}{{\cal O}}

\newcommand{\R}{{\bf R}}
\newcommand{\N}{{\bf N}}
\newcommand{\C}{{\bf C}}

\newcommand{\un}{{\bf I}}

\begin{document}

\centerline{\large\bf Exponential Estimates in Adiabatic Quantum 
Evolution\footnote{ Proceedings of $\mbox{XII}^{\mbox{th}}$
International Congress of Mathematical Physics, Brisbane, Australia,
1997.}}
\vspace{.3cm}
\centerline{\bf \underline{Alain Joye}\footnote{Institut Fourier,
Universit\'e  Grenoble-1, B.P. 74, F-38402 Saint-Martin d'H\`eres} and 
Charles-Edouard Pfister\footnote{
D\'epartement de Math\'ematiques,
Ecole Polytechnique F\'ed\'erale de Lausanne,
CH-1015 Lausanne}}

\vspace{.2cm}
\centerline{\bf Abstract}

{ \small We review recent results concerning the exponential 
behaviour of transition probabilities across a gap in the adiabatic 
limit of the time-dependent Schr\"odinger equation. They range 
from an exponential estimate in quite general situations to 
asymptotic Landau-Zener type formulae for finite dimensional 
systems, or systems reducible to this case. }

\centerline{\bf 1 Introduction}

The notion of adiabatic evolution or adiabatic process is an important theoretical concept, which 
occurs at several places in Physics. In Quantum Mechanics, this process is usually described by the
equation 
$i\hbar{\partial \over\partial t'}\psi(t')=H(\varepsilon t')\psi(t')$,
where  $\varepsilon$ is a small parameter, such that $1/\varepsilon$ gives the typical time-scale over
which the Hamiltonian changes significantly. Setting $\hbar =1$ and introducing a rescaled time,
$t=\varepsilon t'$, we can rewrite the time-dependent Schr\"odinger equation for the evolution operator 
$U$ as
\be\label{sch}
  i\eps\frac{\partial}{\partial t}U(t,s)=H(t)U(t,s)\,, \quad U(s,s)=\un\ ,\quad
  \forall t,s\in \R.
\ee
The adiabatic limit corresponds to the singular limit  $\eps\ra 0$ of the equation (\ref{sch}). 
If the state of the system is an eigenfunction
$\psi(t_0)$ for the eigenvalue $e(t_0)$ at $t=t_0$, then, in the adiabatic limit, the state of the system
at time $t=t_1$ is an eigenfunction $\psi(t_1)$ for the eigenvalue $e(t_1)$, provided the energy-level
$e(t)$ is isolated in the spectrum of the Hamiltonian $H(t)$ for all $t$ in the time-interval
$[t_0,t_1]$ [4].  When the system performs a cycle, $H(t_0)=H(t_1)$, and the energy-level $e(t)$ is
non-degenerate, the eigenfunction $\psi(t_1)$ defined in the adiabatic limit differs from $\psi(t_0)$ by a
phase, which can be decomposed into an $\eps$-dependent dynamical phase and an  
$\eps$-independent geometrical phase related to the spectral subspaces visited during the adiabatic
evolution. This is the fundamental observation of Berry, which gave rise to extensive developments
(see the collection of papers in [32]). In this note we review a complementary aspect, namely 
the estimation of the probability that the final state of the system is {\em not} an eigenstate  
$\psi(t_1)$ for $e(t_1)$; such a transition is called adiabatic or nonadiabatic transition in the
physical literature. There are two kinds of results. On the one hand for small systems 
(or systems reducible to this case), typically two-level systems,  one can derive explicit formulae
for the probability of such a transition.
A particular case is the Landau-Zener formula used when the two levels display an avoided crossing. 
For general systems, on the other hand, one can 
usually obtain 
upper estimates for the transition probability only. Those estimates are reviewed in the second part.
The whole discussion is lead in a scattering setting under the assumption that the
time-dependence of the Hamiltonian is analytic.
More precisely\\
{\bf H1} {\it The self-adjoint family $\{H(t)\}_{t\in\R}$ is defined
on a common dense domain $D$ of a separable Hilbert space ${\cal H}$; it is uniformly bounded from
below; for each $\phi\in D$, $H(t)\phi$ has an analytic extension
in a fixed open strip $S\subset \C$ including the real axis. Moreover,
there exist $H(\pm\infty)$, two self-adjoint operators defined on $D$, and constants 
$C,\alpha>0$ such that for $t{\scriptstyle > \atop <} 0$, 
$\sup_{s\, |\, t+is\in S}\|(H(t+is)-H(\pm\infty))\ffi\|\leq C\frac{\|\ffi\|+\|H(\pm\infty)\ffi\|}
{(1+|t|)^{1+\alpha}}$.}\\
{\bf H2} {\it The spectrum $\sigma(t)$ of $H(t)$ consists in two disjoint parts $\sigma(t)=
\sigma_1(t)\cup \sigma_2(t)$, such that $\inf_{t\in\R}\mbox{dist }(\sigma_1(t),\sigma_2(t))\geq g >0$
and $\sigma_1(t)$ is bounded.}\\ 
Denoting by $P_j(t)$ the spectral projector associated with $\sigma_j(t)$, $j=1,2$, 
the transition probability between the subspaces $P_1(t_0){\cal H}$ et $P_2(t_1){\cal H}$ is defined as
${\cal P}(t_1,t_0,\eps)=\|P_2(t_1)U(t_1,t_0)P_1(t_0)\|^2$.
The  Adiabatic Theorem of Quantum Mechanics  
[4], [21], [28], [1] states that ${\cal P}(t_1,t_0,\eps)=
\ode (\eps^2)$ as soon as $H(t)$ is $C^3$. Under the assumption H1 (analyticity {\em and} existence
of $H(\pm\infty)$) we have ${\cal P}(+\infty,-\infty,\eps)=\ode (\eps^{\infty})$ [25], [29], [22]. 
The difficult questions are to prove exponential estimates and 
asymptotic formulae for ${\cal P}(+\infty,-\infty,\eps)$.

\centerline{\bf 2 Asymptotic formulae}

Assume the Hamiltonian $H(t)$ is an $n\times n$ hermitian matrix with
non degenerate spectrum $\sigma(t)=\{e_1(t),\cdots,e_n(t)\}$. We determine up to a global phase factor
a corresponding basis of normalized eigenvectors $\ffi_j(t)$, $j=1,\cdots,n$, by requiring that 
$\bra \ffi_j(t)|\ffi_j'(t)\ket\equiv 0$, $t\in\R$. 
We can expand the solution $\psi(t)=U(t)\psi(0)$ of (\ref{sch}) as
\be\label{expa}
  \psi(t)=\sum_{j=1}^nc_j(t)\e^{-i\int_0^te_j(t')dt'/\eps }\ffi_j(t),
\ee
where the $c_j's$ are complex valued coefficients and the phase factors
$\e^{-i\int_0^te_j(t')dt'/\eps }$ are introduced for convenience. It follows from H1 that
the limits $\lim_{t\ra\pm\infty}c_j(t)=c_j(\pm\infty)$ exist. Thus, choosing 
$c_j(-\infty)=\delta_{jk}$ as initial conditions, we get that $|c_j(\infty)|^2$ yields the
transition probability from the eigenspace associated with $e_k(-\infty)$ to the 
eigenspace associated with $e_j(+\infty)$, which we denote by ${\cal P}_{jk}(\eps)$. Born and
Fock [4] showed that
\be\label{esre}
  c_j(t)=c_j(-\infty)+\ode (\eps)
\ee
 uniformly in $t\in\R$, from which the Adiabatic Theorem ${\cal P}_{jk}(\eps)=\ode( \eps^2)$ follows.

In case $n=2$, for real symmetric analytic Hamiltonians, the pioneering works [24], [34] and 
[5] gave arguments in favour of the Landau-Zener-Dykhne formula ${\cal P}_{12}(\eps)\simeq
\e^{-2\gamma /\eps}$ with $\gamma >0$ explicited below. A convincing derivation of this formula 
based on the integration of (\ref{sch}) in the complex $t$-plane can be
found in the important contribution [9]. However, for {\it complex} hermitian matrices, 
this formula misses a prefactor of geometrical nature, as we now show. The proof is based on the
fact that the solution $\psi$ of (\ref{sch}) is analytic throughout the strip $S$, since $H$
is, whereas $\ffi_j$ and $e_j$ have multivalued extensions. More precisely, their only possible 
branching points are the complex crossing points $z_0\in S$, such that $e_1(z_0)=e_2(z_0)$. Generically, 
$e_2(z)-e_1(z)\simeq \sqrt{z-z_0}$. Let $\eta_0$ be a negatively oriented
loop based at the origin, which encircles $z_0$. Denoting by $f(z|\eta_0)$ the analytic continuation along 
$\eta_0$ of a function $f(z)$ defined in a neighbourhood of the origin, we can write $e_1(z|\eta_0)=e_2(z)$.
It follows that $\ffi_1(z|\eta_0)$ is proportional to $\ffi_2(z)$ and we define 
$\theta_{21}(\eta_0)\in\C$ by 
$\ffi_1(z|\eta_0)=\e^{-i\theta_{21}(\eta_0)}\ffi_2(z)$. 
Since $\psi(z|\eta_0)=\psi(z)$, we deduce from the foregoing and (\ref{expa}) the key identity 
\be\label{fun}
  c_1(z|\eta_0)=\e^{i\theta_{21}(\eta_0)}\e^{i\int_{\eta_0}e_1(z')dz'/\eps}c_2(z).
\ee

Let us assume, for simplicity, that there exists a unique 
generic crossing point $z_0$ in $S$ with $\mbox{Im }z_0 >0$. By analyticity of $\psi$, we can integrate 
(\ref{sch}) from $-\infty$ to $+\infty$ with 
$ c_j(-\infty)=\delta_{1j}$
along the real axis or along any path $\beta\subset S$ which passes above $z_0$. Denoting by 
$\widetilde{c_1}$ the analytic continuation of $c_1$ along such a path $\beta$, we have, due to
(\ref{fun}), 
$c_2(+\infty)=\e^{-i\theta_{21}(\eta_0)}\e^{-i\int_{\eta_0}e_1(z')dz'/\eps}\widetilde{c_1}(+\infty)$. It
thus remains to prove the estimate $\widetilde{c_1}(+\infty)=1+\ode(\eps)$, which is the equivalent along 
$\beta$ of the estimate (\ref{esre}). As is well known from complex WKB methods, such an estimate can
be proven under a {\it global} dissipativity condition on the path $\beta$ only. By definition, a path 
$\beta\subset S$ 
parametrized by $t\mapsto \beta(t)$ with $\lim_{t\pm\infty}\mbox{Re }\beta(t)=\pm\infty$ is called 
{\it dissipative} (for the indices $\{1,2\}$) if $\mbox{Im }\int_{\beta(t)}(e_1(z)-e_2(z))dz$ is 
non-decreasing in $t\in\R$, where 
$\int_{\beta(t)}$ means integration from $-\infty$ to $\beta(t)$ along 
$\beta(s), -\infty\leq s\leq t$. 
Sumerizing the above discussion we have\\
{\bf Theorem 1 [12]} {\it Assume H1, H2 for $H(t)$ a $2\times 2$ matrix and suppose there 
exists a unique generic crossing point $z_0$ in $S$ with $\mbox{Im }z_0 >0$. Then, provided there exists
a dissipative path in $S$ going from $-\infty$ to $+\infty$ above $z_0$, we have for $\eps >0$ small enough
\be
  {\cal P}_{21}(\eps)=\e^{2\mbox{\scriptsize Im }\theta_{21}(\eta_0)}
  \e^{2\mbox{\scriptsize Im }\int_{\eta_0}e_{1}(z)dz/\eps}(1+\ode(\eps)).
\ee}
\noindent{\bf Remarks:} 
The complex factor $\e^{-i\theta_{21}(\eta_0)}$ is 
actually obtained by analytic continuation along $\eta_0$
of the quantity yielding the geometric phase when considered on the real axis.  When $H(t)$ is real 
symmetric, $\mbox{Im }\theta_{21}(\eta_0)= 0$ and we recover the Landau-Zener-Dykhne formula. However, 
in general, $\mbox{Im }\theta_{21}(\eta_0)\neq 0$, thus giving rise to a non-trivial prefactor. 
This prefactor was independently
discovered by Berry [3].\\
The (non-trivial) questions of existence of dissipative paths and competition between several crossing 
points in $S$ are analyzed in details in [12]. 
\\ In case no dissipative path above $z_0$ exists, an 
exponential bound follows from integration along dissipative paths close to the real axis [12], 
[20]. See also Theorem 2.\\
We refer the reader to tables 1 and 2 in [16] and references therein for similar 
asymptotic formulae in more general $2$-level systems as well as in non-generic situations. 

When $H(t)$ is an $n\times n$ matrix with $n>2$ the same strategy is applicable in principle. 
Let us see this for $n=3$, to fix the ideas, in the following simple setting. Suppose there exist two 
distinct crossing points $z_0$ and $z_1$ only in $S$, with $\mbox{Im }z_j>0$. Let $z_0$ be a generic crossing 
point for the analytic continuations of $e_1(t)$ and $e_2(t)$ around a loop $\eta_0$ 
encircling $z_0$ only, as above. Assume the analytic continuation of $e_3(t)$ in a neighborhood of 
$\eta_0$ is analytic at $z_0$, as is generically the case. 
By hypothesis, $z_1$ is a generic crossing point for the analytic continuations of $e_2(t)$ and $e_3(t)$ 
around a similar loop $\eta_1$ encircling $z_1$ only, whereas the analytic continuation
of $e_1(t)$ in a neighbourhood of $\eta_1$ is analytic at $z_1$. We get, as above, that
(\ref{fun}) holds and must be completed by
\be\label{fun2}
    c_2(z|\eta_1)=\e^{i\theta_{32}(\eta_1)}\e^{i\int_{\eta_1}e_2(z')dz'/\eps}c_3(z),
\ee
where $\theta_{32}(\eta_1)$ is defined similarly. Assume the composition of the two negatively
oriented loops, based at the origin, $\eta_0$ 
followed by $\eta_1$ can be deformed into one negatively oriented loop, based at the origin, encircling both 
$z_0$ and $z_1$. Then, if $\widetilde{c_1}$ denotes the analytic
continuation of $c_1$ along a path going from $-\infty$ to $+\infty$ above $z_0$ and $z_1$, we get from
(\ref{fun}) and (\ref{fun2})
 $c_3(+\infty)=\e^{-i\theta_{21}(\eta_0)}\e^{-i\int_{\eta_0}e_1(z')dz'/\eps}
\e^{-i\theta_{32}(\eta_1)}\e^{-i\int_{\eta_1}e_2(z')dz'/\eps}\widetilde{c_1}(+\infty)$. Again,
it remains to prove $\widetilde{c_1}(+\infty)=1+\ode (\eps)$ with 
initial conditions $ c_j(-\infty)=\delta_{1j}$. Such an estimate can be proven in a dissipative 
domain, defined as follows. A domain $P\subset S$ is {\it dissipative} if it extends from $-\infty$
to $+\infty$ and if any point $z\in P$ can be reached from $-\infty$ by a dissipative path 
$\beta_2\subset P$ for the indices $\{1,2\}$ {\it and} by a dissipative path $\beta_3\subset P$ for the 
indices $\{1,3\}$. However, generally, dissipative domains going above $z_0$ and $z_1$ do not extend 
from $-\infty$ to $+\infty$ [6], [9]. The way out is to perform a perturbative analysis.

Assume $H(t,\delta)$ depends on a supplementary parameter $\delta \geq 0$, satisfies H1 for each 
fixed $\delta$ and is regular enough in $(t,\delta)$. Suppose that for $\delta=0$, the 
(analytic) eigenvalues $e_j(t,\delta=0)$, $t\in\R$, labelled in increasing order at $t=-\infty$, are non 
degenerate, except for two generic real crossings at $t_0<t_1$ where $e_1(t_0,0)=e_2(t_0,0)$ and 
$e_1(t_1,0)=e_3(t_1,0)$. Assume these degeneracies are lifted as soon as $\delta>0$, turning $t_0$ and 
$t_1$ into {\it avoided crossings}. As a consequence,
for $\delta>0$ small enough, there exist crossing 
points $z_0(\delta)$ and $z_1(\delta)$, with the properties discussed above and 
$\lim_{\delta\ra 0}z_j(\delta)=t_j$. Furthermore, the existence of a suitable dissipative domain 
$P$ can be proven perturbatively so that we get 
\\{\bf Theorem 1' [11]} {\it Assume $H(t,\delta)$ is a $3\times 3$ matrix satisfying the hypotheses 
(loosely) stated in the previous paragraph. Then, for $\delta, \eps >0$ small enough,
\be
  {\cal P}_{31}(\eps)=\e^{2\mbox{\scriptsize Im }\theta_{21}(\eta_0,\delta)}\e^{2\mbox{\scriptsize Im }\int_{\eta_0}
e_1(z',\delta)dz'/\eps}
\e^{2\mbox{\scriptsize Im }\theta_{32}(\eta_1,\delta)}\e^{2\mbox{\scriptsize Im }
\int_{\eta_1}e_2(z',\delta)dz'/\eps}
(1+\ode(\eps)),
\ee
where $\ode(\eps)$ is uniform in $\delta$ and $\lim_{\delta\ra 0}\mbox{Im }\theta_{ij}
(\eta_k,\delta)= \lim_{\delta\ra 0}\mbox{Im }\int_{\eta_k}e_j(z',\delta)dz'=0$.
}
{\bf Remarks:} The above result and
generalizations in [11], [18] show that it is possible to get asymptotic formulae for 
{\it certain} transition probabilities in $n$-level systems displaying avoided crossings, despite the 
difficulties induced by the notion of dissipativity.\\ 
The same approach was used when two levels only display one avoided crossing [10]. It yields 
back Theorem 1 without need to check dissipativity conditions in the complex plane. Moreover, assuming
the generic behaviour $e_2(t)-e_1(t)\stackrel{\delta\ra 0}{\simeq}\sqrt{a^2t^2+\delta^2}$,
we get the Landau-Zener formula: $2\mbox{ Im }\int_{\eta_0}
e_1(z',\delta)dz'/\eps =-\frac{\pi\delta^2}{2a\eps}(1+{\cal O} (\delta))$ and 
$\mbox{Im }\theta_{21}(\eta_0,\delta)=\ode (\delta)$ [10].
Further refinements of the Landau-Zener formula can be found in [27].\\
\centerline{\bf 3 Exponential estimates}

Let us come back to general abstract time-dependent Hamiltonians. In such cases we can prove 
that the transition probability across the gap is exponentially small.\\
{\bf Theorem 2 [13]} {\it Let $H(t)$ be a time-dependent Hamiltonian satisfying assumptions 
H1 and H2. Then, there exist $C,\gamma>0$ and 
$\eps_0>0$ such that for all $0<\eps <\eps_0$,
\be {\cal P}(+\infty,-\infty,\eps)=\lim_{t_0\ra -\infty \atop t_1\ra +\infty}
\|P_2(t_1)U(t_1,t_0)P_1(t_0)\|^2 \leq C\e^{-2\gamma/\eps}\ee
}
The proof is made by integrating (\ref{sch}) in the complex plane along 
generalized dissipative paths suitable for operators (see [13], section 5), 
following [12]. A weaker result can be found in [19]. 
Similar estimates were known for 
finite dimensional systems of ODE's, see e.g. [7] and the literature quoted in [6].
Theorem 2 was then recovered by different methods, some of which allowing a better
control on the exponential decay rate $\gamma$ as a function of the gap $g$. See [26] for an 
approach using microlocal analysis, [30] and [15] for superadiabatic techniques, see below, 
and [33] for a pseudo-differential point of view. It follows from  [15]
and [26] that $\gamma \geq c g$, $g$ large, for some constant $c$.

\centerline{\bf 4 Superadiabatic renormalization and reduction theory}

The root of this method is the work [8] and subsequent generalizations and adaptations [29], 
[31], [14] on iterative schemes. Since the work [21], the Adiabatic Theorem is often proven by 
showing that the {\it adiabatic evolution} $V(t,s)$ defined by the equation
\be\label{adiab}
  i\eps\frac{\partial}{\partial t}V(t,s)=(H(t)+i\eps[P_1'(t),P_1(t)])U(t,s)\,, \quad V(s,s)=\un\ ,\quad
  \forall t,s\in \R
\ee
satisfies the intertwining property $V(t,s)P_j(s)=P_j(t)V(t,s)$ 
and $\|U(t,s)-V(t,s)\|=\ode (\eps)$ [1]. Such an approach can be used in a wider context, 
see [23].
An iterative scheme consists in generating from $H(t)$, by standard perturbation in $\eps$, a sequence 
of Hamiltonians $\{ H_q(t,\eps)\}_{q\in\N}$ with
the following features:\\
i) The Hamiltonians $H_q(t,\eps)$ share the same 
general properties as $H(t)$. In particular, the gap hypothesis is true for $H_q(t,\eps)$ so that
the spectral projectors $P_{j,q}(t,\eps)$ are well defined and tend to $P_j(t)$ as $\eps\ra 0$.\\
ii) The adiabatic evolution $V_q(t,s)$, defined by (\ref{adiab}) with $H_q(t,\eps)$ and $P_{j,q}(t,\eps)$ 
in place of $H(t)$ and $P_j(t)$, approximates $U(t,s)$ up to order
$\ode(\eps^q)$ instead of $\ode (\eps)$. 

It follows that
the transition probability between spectral subspaces of $H_q(t,\eps)$ is $\ode (\eps^{2q})$ instead 
of $\ode (\eps^2)$ for finite times. Berry [2] formally showed for two-level systems
that one can push the estimates up to exponential order by truncating the scheme at optimal $q\in\N$. 
In the present setting, Nenciu proved such exponential estimates in [30] using closely related ideas. 
The result of Nenciu was recovered in [15] by optimal truncation of the iterative scheme
proposed in [14].\\
{\bf Theorem 3 [30], [15]} {\it Assume H1 and H2. Then there exist constants 
$C,\gamma,\eps_0 >0$ and a self adjoint 
operator $H_*(t,\eps)$ defined on $D$ such that for all $0<\eps <\eps_0$ and $t\in\R$, 
$\|H_*(t,\eps)-H(t)\|\leq C\eps /(1+|t|)^{(1+\alpha)}$. Moreover, the adiabatic
evolution $V_*(t,s)$ associated with $H_*(t,\eps)$ satisfies
  $\sup_{t,s\in\R}\|V_*(t,s)-U(t,s)\|\leq C\e^{-\gamma/\eps}$ as well as 
$V_*(t,s)P_{j,*}(s,\eps)=P_{j,*}(t,\eps)V_*(t,s)$. Finally, $\gamma\geq C g$, for $g$ large.
}\\
{\bf Remarks:} Theorem 2 becomes a corollary of Theorem 3, since $P_{j,*}(t,\eps)
\stackrel{t\ra\pm\infty}{\ra}P_j(\pm\infty)$. 
The evolution $V_*(t,s)$ is called {\it superadiabatic evolution} due to the 
exponential estimate. 

 The main interest of this construction, however,
is that it allows to set up a rigorous reduction theory. Assume $P_1(t){\cal H}$ is finite dimensional,
say two-dimensional, such that $\sigma_1(t)=\{e_1(t),e_2(t)\}$ where $e_j(t)$ satisfy the gap assumption.
Then the transition probability ${\cal P}_{21}(\eps)$ between eigenspaces associated with $e_1(-\infty)$
and $e_2(+\infty)$ can be computed modulo errors $\ode (\e^{-\gamma/\eps})$ by replacing the evolution 
$U(t,s)$ by the superadiabatic evolution $V_*(t,s)$ which describes the evolution {\it inside } the 
$P_{j,*}(t,\eps){\cal H}$. This leads to an {\it effective} two-dimensional problem, with a corresponding 
effective $2\times 2$ Hamiltonian shown to be a perturbation of $H(t)P_1(t)$ 
 and to have the same analyticity properties as $H(t)$ in $S$. Hence, provided the conditions on the
analytic continuations of $e_j$ stated
in Theorem 1 are satisfied, we get with the same notations [15]
\be
  {\cal P}_{21}(\eps)=\e^{2\mbox{\scriptsize Im}\theta_{21}(\eta_0)}
\e^{2\mbox{\scriptsize Im}\int_{\eta_0}e_1(z)dz/\eps}(1+\ode(\eps))+\ode(\e^{-\gamma/\eps})
\ee
The condition
$\gamma > |2\mbox{Im }\int_{\eta_0}e_1(z)dz|$ is satisfied if $g$ is large enough [15] or in the
avoided crossing regime described in Theorem 1' [10], [11].

Another use of the superadiabatic renormalization consists in performing the analysis of 
section 2 for finite dimensional systems when the decomposition (\ref{expa}) 
is replaced by
  $\psi(t)=\sum_{j=1}^nc_{j,*}(t)\e^{-i\int_0^te_{j,*}(t',\eps)dt'/\eps }\ffi_{j,*}(t,\eps)$
where $e_{j,*}(t,\eps)$ and $\ffi_{j,*}(t,\eps)$ are the eigenvalues and eigenvectors of $H_*(t,\eps)$.
After showing that they have analytic continuations in $S$ with similar properties as those of $e_j$ and
$\ffi_j$, we get very accurate asymptotic formulae by making use of 
Theorem 3 "in the complex plane". Under the conditions of Theorem~1  
\be
  {\cal P}_{21}(\eps)=\e^{2\mbox{\scriptsize Im}\theta_{21,*}(\eps)}
\e^{2\mbox{\scriptsize Im}\int_{\eta_0}e_{1,*}(z,\eps)dz/\eps}(1+\ode(\e^{-\gamma/\eps}))), \,\,\,
\gamma >0,
\ee 
as shown in [17], [11]. The loop $\eta_0$ and $\theta_{21,*}(\eps)$ are defined as in section 2.\\
\centerline{\bf References:}
[1] J.E. Avron, R. Seiler, L.G. Yaffe: {\em Commun.Math.Phys}{\bf 110} (1987)
33-49.
\newline
[2] M.V. Berry:
            {\em Proc.Roy.Soc.Lond.A} {\bf 429} (1990) 61-72.
\newline
[3] M.V. Berry: {\em Proc.Roy.Soc.London A}{\bf 430} (1990) 405-411.
\newline
[4] M. Born, V. Fock: {\em Zeit.f.Physik}
            {\bf 51} (1928) 165-180.
\newline
[5] A.M. Dykhne:
        {\em Sov.Phys.JETP.} {\bf 14} (1962) 941-943.
\newline
[6] M. Fedoriuk:   "M\'ethodes Asymptotiques pour les Equations 
Diff\'erentielles Ordinaires Lin\'eaires", Mir Moscou 1987.
\newline
[7] N. Fr\"oman, P.O. Fr\"oman,  " JWKB Approximation,
              Contributions to the Theory", North Holland 1965.
\newline
[8] L.M. Garrido: {\em J.Math.Phys.}
{\bf 5} (1964) 335-362.
\newline
[9] J.-T. Hwang, P. Pechukas: {\em J.Chem.Phys.}{\bf 67} (1977) 4640-4653.
\newline
[10]  A. Joye:
{\it    Asymp. Anal.} {\bf  9} (1994) 209-258.
\newline
[11] A. Joye:
{\it SIAM J. Math. Anal.} {\bf 28} (1997) 669-703.
\newline
[12] A. Joye, H. Kunz, C.-E. Pfister: {\em Ann.Phys.} {\bf 208} (1991) 299-332.
\newline
[13] A. Joye, C.-E. Pfister:
{\em Commun.Math.Phys.}{\bf 140} (1991) 15-41.
\newline
[14] A. Joye, C.-E. Pfister: {\em J.Phys.A}{\bf 24} (1991) 753-766.
\newline
[15] A. Joye, C.-E. Pfister:
{\em J.Math.Phys.}{\bf 34} (1993) 454-479.
\newline
[16] A. Joye, C.-E. Pfister in
NATO ASI Series B: Physics {\bf 324} Plenum, New York, (1994) 139-148.
\newline
[17] A. Joye, C.-E. Pfister : 
        {\it SIAM J. Math. Anal.} {\bf 26} (1995) 944-977.
\newline
[18] A. Joye and C.-E.Pfister : 
{\it CPT-Marseille Preprint} (1997).
\newline
[19] V. Jaksic, J. Segert: {\em Rev.Math.Phys.}{\bf 4} (1992) 529-574.
\newline
[20] V. Jaksic, J. Segert: {\em J.Math.Phys.}{\bf 34} (1993) 2807-2820.
\newline
[21] T. Kato:
{\em J.Phys.Soc.Japan} {\bf 5} (1950) 435-439.
\newline
[22] M. Klein, R. Seiler: {\em Commun.Math.Phys.}{\bf 128} (1990) 141-160.
\newline
[23] S.G. Krein, "Linear Differential Equations in Banach Space",
            AMS, Providence 1971.
\newline
[24] L.D. Landau: {\em Collected Papers of L.D.Landau}, Pergamon Press,
Oxford London Edinburgh New York Paris Frankfurt 1965.
\newline
[25] A. Lenard: Adiabatic Invariance to all Orders,
{\em Ann.Phys} {\bf 6} (1959) 261-276.
\newline
[26] A. Martinez:
{\em J.Math.Phys.} {\bf 35} (1994) 3889-3915.
\newline
[27] P.A. Martin, G. Nenciu: {\em Rev.Math.Phys.} {\bf 7}
(1995) 193-242.
\newline
[28] G. Nenciu:
{\em J.Phys.A}{\bf 13} (1980) L15-L18.
\newline
[29] G. Nenciu:
{\em Commun.Math.Phys.}{\bf 82} (1981) 121-135.
\newline
[30] G. Nenciu: {\em Commun.Math.Phys.}{\bf 152} (1993) 479-512 
\newline
[31] G. Nenciu, G. Rasche: {\em Helv.Phys.Acta}{\bf 62} (1989) 372-388.
\newline
[32] A. Shapere, F. Wilczek: "Geometric Phases in Physics", World Scientific,
Singapore, New Jersey, London, Hong Kong 1989.
\newline
[33] J. Sj\"ostrand:
{\em C.R.Acad.Sci.Paris} t.317  S\'er.I {\bf 22} (1993) 217-220.
\newline
[34] C. Zener:
            {\em Proc.Roy.Soc.London}{\bf 137} (1932) 696-702.

\end{document}